\documentclass[11pt, a4paper]{article}
\pdfoutput=1
\usepackage[utf8]{inputenc}
\usepackage{jheppub}
\usepackage{amsmath}
\usepackage{graphicx}
\usepackage{stix}

\DeclareMathOperator{\Tr}{Tr}
\DeclareMathOperator{\dimension}{dim}

\usepackage{natbib}
\setcitestyle{square, comma, numbers,sort&compress}

\title{Scalar Insertions in Cusped Wilson Loops in the Ladders Limit of Planar \boldmath{$\mathcal{N}=4$} SYM}

\author[]{Joseph McGovern }

\affiliation[]{Mathematics Department, King's College London, The Strand, London WC2R 2LS, UK\\
Mathematical Institute, University of Oxford, Radcliffe Observatory, OX2 6GG, UK}

\emailAdd{mcgovernjv@gmail.com}

\abstract{Compact expressions in terms of the Q-functions of the Quantum Spectral Curve are given for 3-cusped fundamental Wilson loops in the ladders limit of $\mathcal{N}=4$ Super Yang-Mills with additional scalars inserted at a cusp between uncoupled arcs. This gives some further credence to the already natural and evidenced view that the Quantum Spectral Curve in fact pertains to a wide class of observables beyond the spectrum, as well as providing additional nonperturbative ladders limit results.}

\begin{document}

\maketitle
\flushbottom

\section{Introduction}

The integrability approach to planar $\mathcal{N}=4$ Super Yang-Mills excitingly allows for non-perturbative results in this interacting field theory --- see \cite{Gromov:2017blm} for a pedagogical introduction and \cite{Beisert:2010jr} for a wider review. While the full theory remains unsolved, many advances have been made towards this end. The Thermodynamic Bethe Ansatz allowed for the cusp anomalous dimension to be determined in \cite{Correa:2012hh} and \cite{Drukker:2012de} after the full spectrum of anomalous dimensions for local operators was obtained in \cite{Gromov:2009bc}, \cite{Arutyunov:2009ur} and \cite{Bombardelli:2009ns}. 
In \cite{Basso:2015eqa} the powerful hexagon bootstrap approach was investigated for the computation of correlation functions. One would hope that the Separation of Variables approach would allow access to full non-perturbative expressions for the structure constants in terms of the Q-functions which are solutions of the Quantum Spectral Curve equations. Some examples of non-perturbative correlation functions have been found in \cite{Cavaglia:2018lxi}, \cite{Giombi:2018qox}, \cite{Giombi:2018hsx} and \cite{Kim:2017sju}. Recently, \cite{Jiang:2019xdz} carried out further non-perturbative work for correlation functions of specific operators.
\\

The Quantum Spectral Curve was introduced in \cite{Gromov:2014caa}, where it was used to produce the spectrum of anomalous dimensions for all local single trace operators in the planar $\mathcal{N}=4$ theory. These methods were extended in \cite{Gromov:2015dfa} where they were used to produce the cusp anomalous dimension. Among its numerous strengths the QSC enjoys a relatively simple structure compared to the TBA equations and is highly suggestive from a holographic point of view.
\\

Further progress was made in \cite{Cavaglia:2018lxi}, where for the first time expressions in terms of the Q-functions were given for the structure constants associated to three-cusped Wilson loops in a particular double scaling limit of the theory. These expressions both provided a significant simplification and met the hope that the QSC approach could provide non-perturbative information on many more observables than solely the spectrum. The limit in question is of note for the fact that the only Feynman diagrams that survive the limiting procedure are those consisting of loops containing an arbitrary number of non-crossing scalar propagators. The appearance of these diagrams gives rise to this limiting theory's name, "ladders limit".
\\

This limit was first fully defined in \cite{Correa:2012nk}, but the relevant subset of ladder diagrams was studied in \cite{Erickson:1999qv}. Exact computation was made possible by this relatively simple diagram structure, which allowed for a complete resummation through a Bethe-Salpeter equation. In this setup the Q-function's relevance was manifested in \cite{Cavaglia:2018lxi} through a concise integral transform relating them to the eigenfunctions of the Schr{\" o}dinger operator appearing in this resumming equation. \cite{Cavaglia:2018lxi} studied a similar set of correlators to those in \cite{Kim:2017sju}, and the two sets of work agreed where they overlapped.
\\

In this paper we use the methods of \cite{Cavaglia:2018lxi} to supplement their results with an additional set of exact structure constants. Specifically, we consider objects formed by concatenating three coplanar Wilson arcs and inserting additional scalars at one of the cusps. There are restrictions on the scalar combinations entering on the arcs and at the cusp which we motivate and make clear in the following subsection after introducing necessary terminology. This object is a generalisation of previously considered correlators and a hope is that knowledge of it will provide a starting point for accessing the most general HHH case in which the combinations of scalars attached to each arc is completely arbitrary.
\\

In short, our calculation boils down to identifying that when using the parametrisation of \cite{Cavaglia:2018lxi} every integral coming from each additional scalar propagator can be immediately carried out. This is because the propagator expression is the derivative of a generating function $w_{\phi}$ that appears exponentiated in the aforementioned integral transform formula relating the Q-functions to solutions of the Bethe-Salpeter equation that resums the ladders. We expand on the various details of this statement throughout the paper.
\\

\subsection{Setup}

The Maldacena-Wilson lines in $\mathcal{N}=4$ SYM are given by 

\begin{equation}
\label{line}
W_{x}^{y}\left(\vec{n}\right)=\text{Pexp}\int_{x}^{y}\left(iA_{\mu}dx^{\mu}+\Phi^{a}n^{a}|dx|\right),
\end{equation}

in which $A^{\mu}$ denotes the gauge field, $\Phi^{a}$ $\left(a=1,...,6\right)$ denotes the 6 scalars and $\vec{n}$ is a unit 6-vector. We are studying the theory with gauge group $SU(N)$ in the large $N$ limit. Our attention is restricted to the case where the path of integration in (\ref{line}) is a circular arc (such arcs being related to straight lines by conformal transformations, a symmetry retained in the ladders limit).
\\

\begin{figure}
    \centering
    \includegraphics[width=12cm]{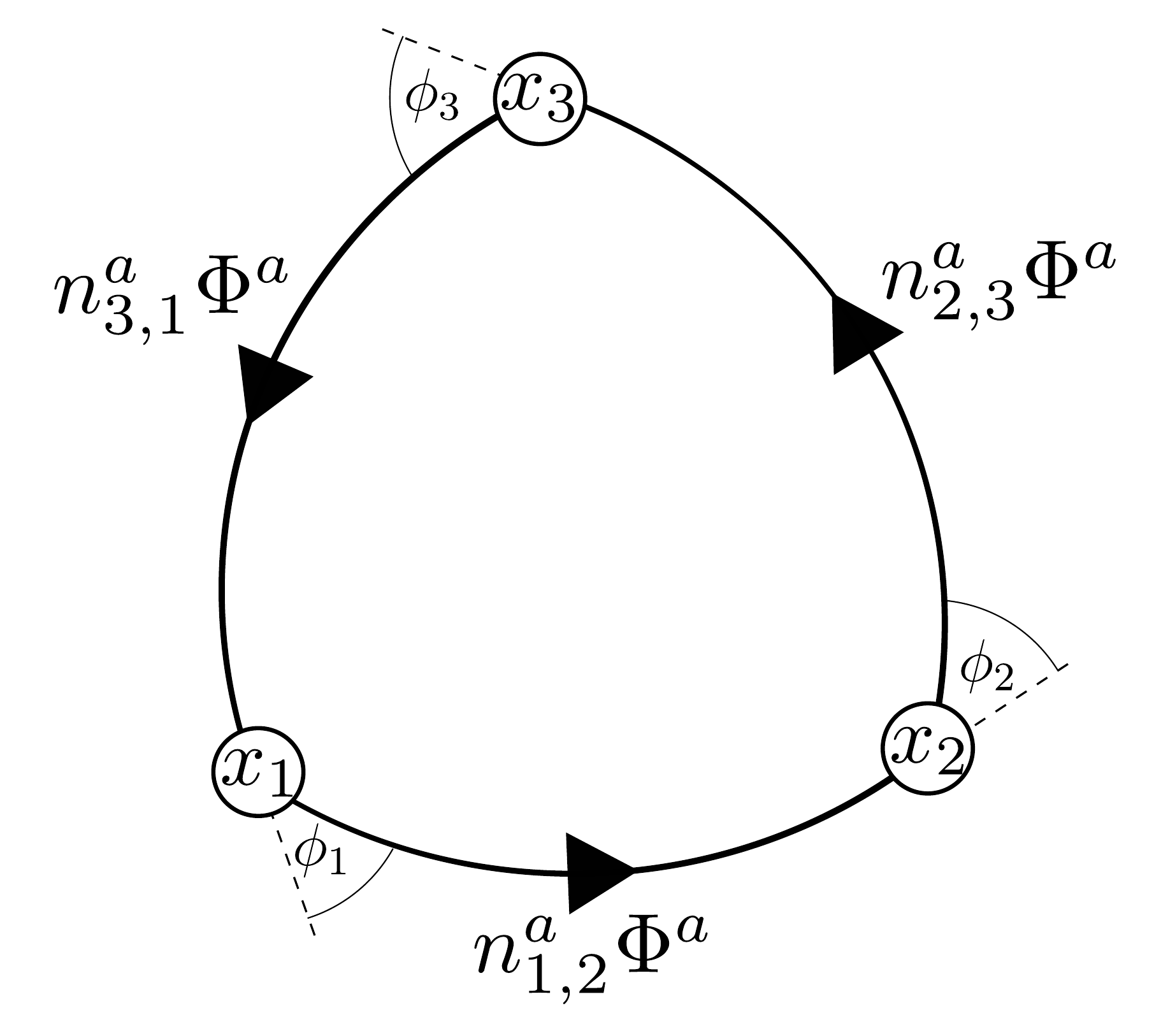}
    \caption{The Wilson loop consisting of three circular arcs, with each arc's respective field coupling indicated.}
    \label{fig:loop}
\end{figure}

The Wilson loop is then formed by concatenating three of these in a two dimensional plane as in figure \ref{fig:loop}. The expectation value of its trace is then taken:

\begin{equation}
\label{loop}
\langle \frac{1}{N}\Tr W_{x_{1}}^{x_{2}}\left(\Vec{n}_{1,2}\right)W_{x_{2}}^{x_{3}}\left(\Vec{n}_{2,3}\right)W_{x_{3}}^{x_{1}}\left(\Vec{n}_{3,1}\right) \rangle.
\end{equation}
\\
This quantity is parametrised by the three angles made at the intersections of the arcs $\phi_{i}$ and the value of each of the inner products $\Vec{n}_{i-1,i}\cdot\Vec{n}_{i,i+1}=\cos\left(\theta_{i}\right)$. It has the form of a 3 point correlator, and it is the structure constant thereof that is of interest. Analysis is restricted to the cases where the angles $\phi_{i}$ obey the triangle inequalities $\phi_{1}+\phi_{2}>\phi_{3}$, $\phi_{2}+\phi_{3}>\phi_{1}$, $\phi_{3}+\phi_{1}>\phi_{2}$ and additionally $0<\phi_{i}<\pi$. Taken together these inequalities fix attention to cases where the intersections of the extensions of any two of the arcs lie outside of the cusped loop. The $\frac{1}{N}$ prefactor is used for loops in the fundamental representation, for a general representation $\mathcal{R}$ one has a factor of $\frac{1}{\dimension(\mathcal{R})}$.
\\

The ladders limit is a double scaling limit in which the 't Hooft coupling $g=\frac{g_{SYM}\sqrt{N}}{4\pi}$ goes to zero with each $\hat{g}_{i}=\frac{g}{2}e^{-i\theta_{i}/2}$ kept finite or zero. These $\hat{g}_{i}$ then provide us with three effective couplings. The sole Feynman diagrams relevant to the Wilson loops that survive this limit are planar ones in which pairs of arcs are connected by scalar propagators (or no propagators). It is in this limit that a concise Bethe-Salpeter equation can be written to resum the ladders (i.e. the scalar propagators between arcs of which there are diagrams containing arbitrarily many, see figure \ref{2cusp}). 
\\

It is worth explicitly mentioning a key property of the scalar propagator's form: $\langle\Tr\left(\Phi^{a}\Phi^{b}\right)\rangle$ is proportional to $g^2\delta^{ab}$, with $g$ the 't Hooft coupling. Therefore for 6-vectors $\vec{r}$ and $\vec{k}$ we have that $\langle\Tr\left(r^{a}\Phi^{a} k^{b}\Phi^{b} \right)\rangle$ is proportional to $g^{2}\delta^{ab}r^{a}k^{b}=g^{2}\vec{r}\cdot\vec{k}$.  
\\

This ladders limit having been taken, one can see that a scalar propagator between the arcs meeting at $x_{i}$ comes with a factor of $\hat{g}_{i}^{2}$. There are then three cases, according to how many of these are nonzero:
\\

$\circ$HLL, where only one $\hat{g}_{i}$ is nonzero.
\\

$\circ$HHL, where exactly two $\hat{g}_{i}$ are nonzero.
\\

$\circ$HHH, the most general case where the three $\hat{g}_{i}$ are arbitrary.
\\

In the above H and L are respectively read as "Heavy" and "Light". One has that a Light cusp is one between arcs not connected by scalar propagators in the Feynman diagrams. This suggestive terminology is justified by the fact that the scaling dimensions of the cusps becomes large at strong coupling (see \cite{Cavaglia:2018lxi}), in agreement with the similar terminology for local operators and the relative sizes of their scaling dimensions.
\\

\subsection{Main results}

Our main result is a set of expressions for HHL Wilson loops with an arbitrary number of additional scalar insertions at the L cusp (the cusp between uncoupled arcs). Written in full this is

\begin{equation}
\label{insertions}
\langle \frac{1}{N}\Tr W_{x_{1}}^{x_{2}}\left(\Vec{n}_{1,2}\right)W_{x_{2}}^{x_{3}}\left(\Vec{n}_{2,3}\right)\left(n_{2,3}^{b}\Phi^{b}\left(x_{3}\right)\right)^{m_{2}}\left(n_{3,1}^{a}\Phi^{a}\left(x_{3}\right)\right)^{m_{1}}W_{x_{3}}^{x_{1}}\left(\Vec{n}_{3,1}\right) \rangle,
\end{equation}

with $\Vec{n}_{2,3}\cdot\Vec{n}_{3,1}=0$. The restriction on the combinations of scalars in the insertions is made for two reasons --- one is that this retains the simple ladder structure of the diagrams (avoiding crossings between "rungs" of the ladders) and the other is that these combinations are those produced by acting on the cusp with \cite{Cavaglia:2018lxi}'s projection operators. We elucidate on the details of this latter statement in section \ref{perspectives}, and for now press on with this restriction applied. 
\\

We find that (\ref{insertions}) can be computed exactly. Ladder resummation about each cusp produces a divergence which is canonically normalised upon division by factors given in section \ref{normal}. These factors were found in \cite{Cavaglia:2018lxi} by studying the loop with two cusps. Factors of $\hat{g}_{1}$ and $\hat{g}_{2}$ are left present in our expression --- one is free to adopt a different normalisation in order to modify these, but such measures would not affect the factors that we focus on here. All dependence on the $\phi_{i}$ is captured by the other factors, as is all nontrivial $\theta_{i}$ dependence. Subject to this normalisation, we find that (\ref{insertions}) equals

\begin{equation}
\label{mainresult}
\hat{g}_{1}^{2m_{1}}\cdot\hat{g}_{2}^{2m_{2}}\cdot\hat{\mathbf{C}}\cdot\left(\frac{|x_{1}-x_{2}|}{|x_{1}-x_{3}||x_{2}-x_{3}|}\right)^{m_{1}+m_{2}}\cdot\left(\frac{|x_{1}-x_{3}|}{|x_{2}-x_{3}||x_{1}-x_{2}|}\right)^{\Delta_{0}^{2}}\cdot\left(\frac{|x_{2}-x_{3}|}{|x_{1}-x_{2}||x_{1}-x_{3}|}\right)^{\Delta_{0}^{1}}.
\end{equation}
\\

In this formula $\hat{\mathbf{C}}$ is a term independent of the $x_{i}$, depending only on the angles $\phi_{i}$. We give a concise expression for $\hat{\mathbf{C}}$ in terms of Q-functions below. Note the correct dependence on $x_{1}$, $x_{2}$ and $x_{3}$ for a three-point correlator in a conformal theory. The dimensions $\Delta_{0}^{i}$ are the same as those given in \cite{Cavaglia:2018lxi}- indeed our result agrees with theirs when $m_{1}$
 and $m_{2}$ are taken to zero.
 \\

We adopt the shorthand $n=m_{1}+m_{2}$, which is the only combination of these numbers relevant outside of the coupling factors. The structure constants $\hat{\mathbf{C}}$ of these quantities are found to have a very simple form when expressed in terms of the Q-functions and a succinct bracket $\langle\; \circ\; \rangle$ to be defined:

\begin{equation}
\label{result}
\scalebox{1.5}{\boxed{\hat{\mathbf{C}}=\frac{\langle q_{0}^{1}\left(u\right)q_{0}^{2}\left(u\right)\frac{e^{-\phi_{3}u}}{u^{n}}\rangle}{\sqrt{\langle\left(q_{0}^{1}\right)^{2} \rangle}\sqrt{\langle\left(q_{0}^{2}\right)^{2} \rangle}} .}}
\end{equation}
\\

The symbol $q_{0}^{i}$ denotes the Q-function associated to cusp $i$ (the subscript 0 refers to a ground state, to be defined precisely in section \ref{section2} but included here for definiteness). The bracket is defined for functions $f(u)\sim u^{\alpha}e^{\beta u}$ at large $u$ by 

\begin{equation}
\label{bracket}
\langle f\left(u\right)\rangle\equiv\left(2\sin\left(\frac{\beta}{2}\right)\right)^{\alpha}\int_{|}f\left(u\right)\frac{du}{2\pi i u},
\end{equation}

in which | is a vertical contour of constant positive real part. The $n=0$ expression is precisely that obtained for the HHL cusped loop in \cite{Cavaglia:2018lxi}, of which (\ref{result}) constitutes a generalisation.
\\

The suggestive use of a bracket to denote the functional (\ref{bracket}) is motivated by the fact that as demonstrated in \cite{Cavaglia:2018lxi} Q-functions of different energies are orthogonal under the bracket. The Q-functions serve as the wave functions in the Separation of Variables approach, and this bracket is then the natural scalar product.
\\

Our results hold for the loop in the fundamental representation. This is ensured by the usual argument that in the large $N$ limit nonplanar diagrams are subleading compared to planar ones and accordingly one does not need to consider diagrams where rungs of the ladders cross. Our results do not hold for arbitrary representations because in general diagrams containing crossed rungs are not subleading for large $N$. For our analysis (which requires planar diagrams) to hold in an arbitrary representation $\mathcal{R}$ then one necessary condition on the generators $T^{A}$ is that as $N\rightarrow \infty$,

\begin{equation}
\label{repstuff}
\frac{\Tr\left( T^{A}T^{B}T^{A}T^{B} \right)}{\Tr\left( T^{A}T^{A}T^{A}T^{B} \right)}=1-\frac{N}{C_{2}\left(\mathcal{R}\right)}+\left(\frac{C_{1}\left(\mathcal{R}\right)}{C_{2}\left(\mathcal{R}\right)}\right)^{2} \rightarrow 0.
\end{equation}
\\

The traces in the above ratio arise from ladder diagrams with two rungs --- the trace in the numerator is produced when the rungs cross and the trace in the denominator is produced when they do not. The formula for this ratio in terms of the casimirs $C_{1}\left(\mathcal{R}\right)$ and $C_{2}\left(\mathcal{R}\right)$ is taken from appendix A of \cite{Correa:2015wma}. This demonstrates for example that in the adjoint, rank-$k$ antisymmetric and rank-$k$ symmetric ($k>1$) representations nonplanar diagrams are not subleading and our analysis fails. Our analysis however is completely safe from this problem when the loops are in the fundamental representation, where in particular the condition (\ref{repstuff}) is met.
\\

For a single scalar insertion the diagrams considered involve a single propagator between the L cusp and the opposite arc on top of the ladder contributions. This propagator is integrated over the arc. In one possible calculation of $\frac{\partial \Delta}{\delta\hat{g}^2}$ for the two-cusped loop a very similar problem is met to which our analysis presents a solution. In section \ref{derivativesection} we detail this calculation and find agreement with a result of \cite{Cavaglia:2018lxi}.

\section{Review of informing material}

\subsection{Q-function machinery and ladder resummation}\label{section2}

\begin{figure}
    \centering
    \includegraphics[width=14cm]{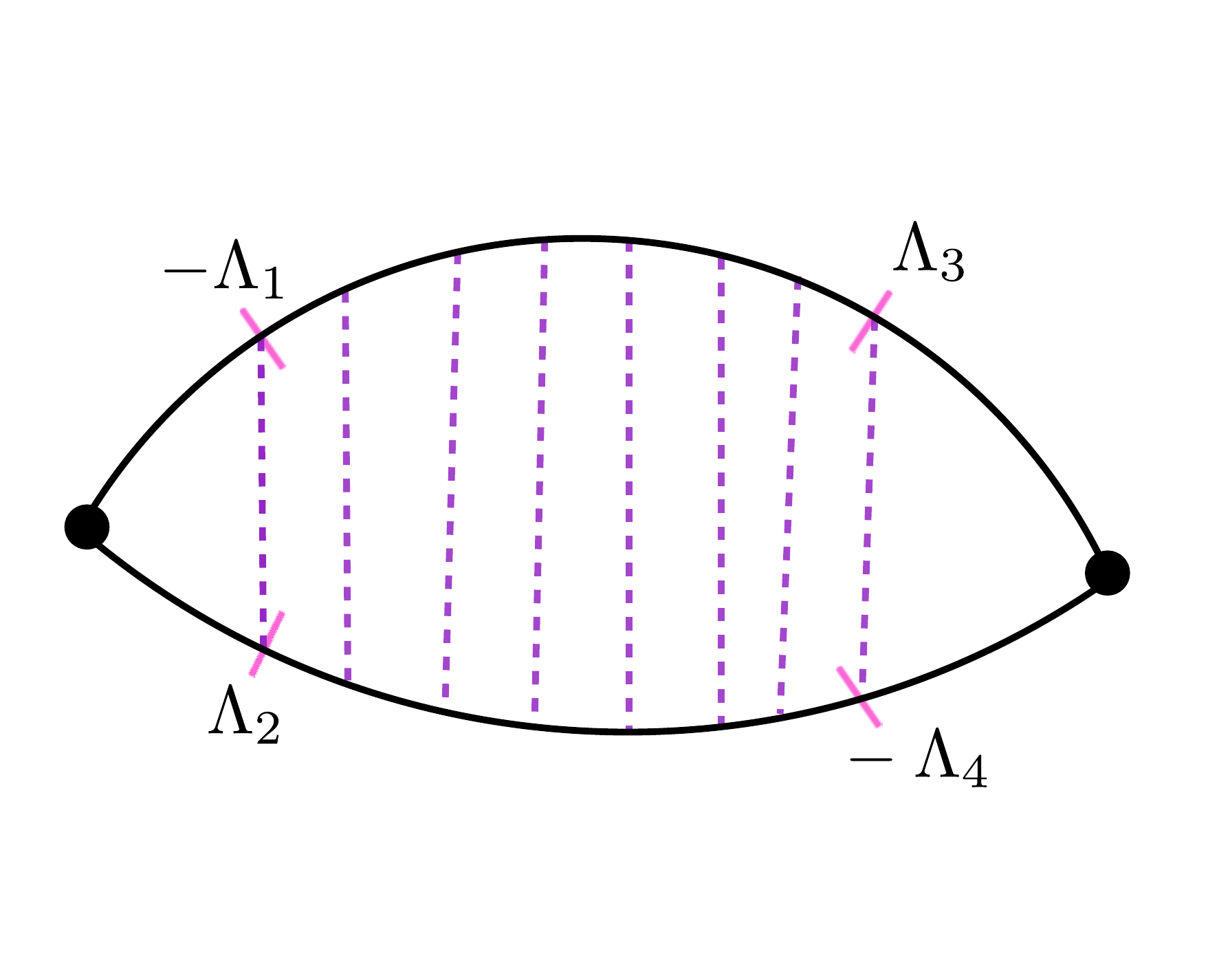}
    \caption{Ladder resummation over part of the 2-cusped loop, defining $G(\Lambda_{1},\Lambda_{2},\Lambda_{3},\Lambda_{4})$. Note that the expression for this is divergent for resummation up to either of the cusps.}
    \label{2cusp}
\end{figure}

The appearance of the Q-functions here is explicitly realised through a relation they share with terms appearing in solutions $G$ of the Bethe-Salpeter equation that resums the ladders for the loops with two cusps (see figure \ref{2cusp}):

\begin{equation}
\label{sum}
G\left(\Lambda_{1},\Lambda_{2},\Lambda_{3},\Lambda_{4}\right)=\sumint_{k}\frac{4F_{k}\left(\Lambda_{1}-\Lambda_{2}\right)F_{k}\left(\Lambda_{4}-\Lambda_{3}\right)}{||F_{k}||^{2}\sqrt{-E_{k}}}\sinh\left(\frac{\sqrt{-E_{k}}}{2}\left(\Lambda_{1}+\Lambda_{2}+\Lambda_{3}+\Lambda_{4}\right)\right).
\end{equation}
\\

The symbol $\sumint_{k}$ denotes summation over the discrete spectrum (here finitely many negative numbers) of a Schr{\" o}dinger problem plus integration over the continuous spectrum (in this case the positive real axis). 
\\

We take the result (\ref{sum}) from \cite{Cavaglia:2018lxi}. A different approach based on an integro-differential equation for resumming ladders was investigated in \cite{Bykov:2012sc}. 
\\

In \cite{Cavaglia:2018lxi} the Bethe-Salpeter equation in question decouples in suitable variables into two equations, one having exponentials as solutions and the other being a Schr{\"o}dinger equation with potential term equal to the scalar propagator. This Schr{\"o}dinger equation was first derived in \cite{Erickson:1999qv}, and we do not write or describe it more fully here as it is not directly useful for the following material. This latter equation possesses a discrete and continuous spectrum of eigenvalues $E_{k}$ with eigenfunctions $F_{k}$, which is what is being summed/integrated over. For small $\hat{g}$ there exists a single bound state having energy in the discrete spectrum, and the number of such states was found in \cite{Cavaglia:2018lxi} to grow linearly with the coupling..
\\

Writing $\Delta_{k}=-\sqrt{-E_{k}}$, the Q-functions's explicit relations to each $F_{k}$ were first given in \cite{Cavaglia:2018lxi} through an integral transform:

\begin{equation}
\label{integraltransform}
F_{k}\left(z\right)=e^{-\Delta_{k}z/2}\int_{|}q_{k}\left(u\right)e^{w_{\phi}\left(z\right)u}\frac{du}{i u}.
\end{equation}

Here the function $w_{\phi}$, which is central to our calculation, is given by

\begin{equation}
\label{w}
e^{iw_{\phi}\left(z\right)}=\frac{\cosh\left(\frac{z-i\phi}{2}\right)}{\cosh\left(\frac{z+i\phi}{2}\right)}.
\end{equation}

The symbol $q_{k}$ denotes the solution to the Quantum Spectral Curve equations at $k^{th}$ excitation. Specifically, the Baxter equation which the Q-functions solve possesses several other solutions for different energy levels. Their relevance to the cusped Wilson loop is manifested in \cite{Cavaglia:2018lxi}, where they enter into the structure constants for cusped loops acted on by exciting projection operators.
\\

The relation (\ref{integraltransform}) is invertible, and the Q-functions formed in this manner indeed obey the Baxter equation coming from the Quantum Spectral Curve formalism (this calculation is carried out in full in \cite{Cavaglia:2018lxi}). Specifically, they are the solutions with large u asymptotics $q(u)\sim u^\Delta e^{\phi u}$.

\subsection{Parametrisation}

Before carrying out our calculation we must also review the parametrisation of \cite{Cavaglia:2018lxi}. This involves an explicit coordinatisation of the arcs with useful highlighted properties and a transfer function for changing direction on the arcs.
\\

We adopt the complex coordinates $x_{i}=(\Re(z_{i}),\Im(z_{i}),0,0)$ and the notation $z_{ij}=|z_{i}-z_{j}|$. The directed arcs connecting $x_{1}=\left(\Re\left(z_{1}\right),\Im\left(z_{1}\right),0,0\right)$ to $x_{2}$ and $x_{3}$ respectively are then the images of  

\begin{equation}
\label{arcs}
\begin{aligned}[b]
\zeta_{13}\left(t\right)&=z_{1}-\frac{z_{12}z_{13}e^{t}}{e^{t}z_{12}+\frac{i}{2\sin\left(\phi_{1}\right)}z_{23}\left(1-e^{t}\right)\left(-e^{-i\phi_{1}}+e^{i\left(\phi_{2}-\phi_{3}\right)}\right)},\\
\zeta_{12}\left(s\right)&=z_{1}-\frac{z_{12}z_{13}e^s}{e^{s}z_{13}+\frac{1}{2\sin\left(\phi_{1}\right)}z_{23}\left(1-e^{s}\right)\left(-e^{i\phi_{1}}+e^{i\left(\phi_{2}-\phi_{3}\right)}\right)}.
\end{aligned}
\end{equation}
\\

Directed arcs connecting $x_{a}$ and $x_{b}$ are in the same way the images of the curves\\
$\vec{x}_{ab}= \left(\Re(\zeta_{ab}(k),\Im(\zeta_{ab}(k) ,0,0\right)$, $k\in (-\infty,0]$. $\zeta_{ab}$ is obtained from (\ref{arcs}) via cyclic permutation of the indices. The principal merit of this parametrisation is how simple the propagator is:
\\

\begin{equation}
\label{derivatives}
\frac{|\dot{\Vec{x}}_{12}\left(s\right)||\dot{\Vec{x}}_{13}\left(t\right)|}{|\Vec{x}_{12}\left(s\right)-\Vec{x}_{13}\left(t\right)|^2}=\frac{1/2}{\cosh\left(s-t-\delta x_{1}\right)+\cos\left(\phi_{1}\right)},
\end{equation}\

\begin{equation}\label{dx1}
\delta x_{1}=\log\left(\frac{\sin\left(\frac{\phi_{1}-\phi_{2}+\phi_{3}}{2}\right)}{\sin\left(\frac{\phi_{1}+\phi_{2}-\phi_{3}}{2}\right)}\right).
\end{equation}
\\

In \cite{Cavaglia:2018lxi}, this particular expression necessarily had the same form as the potential in the resumming Schr{\"o}dinger equation, which allowed for neat cancellations to take place in their deriving of the Q-function expression for the HHL correlator. In our calculation these propagator factors appear once under each integral and handily are the derivative of the function $w_{\phi}$ appearing in the relation (\ref{integraltransform}), allowing the integrals to be straightforwardly carried out. This is essentially all that there is to our calculation.
\\

Since we will be resumming ladders in both directions along the $x_{1}$ to $x_{2}$ arc there is need for a function that relates the two different parametrisations ($\vec{x}_{12}$ and $\vec{x}_{21}$) for a given point on the arc, as in $\zeta_{12}\left(s\right)=\zeta_{21}\left(T_{12}\left(s\right)\right)$. Such a (surprisingly simple) function is given by

\begin{equation}
\label{transfer}
e^{T_{12}\left(s\right)}=\frac{\left(1-e^s\right)}{1-e^{s}\frac{\cos\left(\phi_{3}\right)-\cos\left(\phi_{1}+\phi_{2}\right)}{\cos\left(\phi_{3}\right)-\cos\left(\phi_{1}-\phi_{2}\right)}}.
\end{equation}
\\

\subsection{Normalisation}\label{normal}

The last thing to mention is  the normalisation, which again is the same as in \cite{Cavaglia:2018lxi}. There is a divergence when the ladders around a cusp are resummed, which we normalise by cutting an $\epsilon$-circle around each cusp. This amounts to having $s_{i}$ and $t_{i}$ run from $-\Lambda_{s_{i}}$ and $-\Lambda_{t_{i}}$ respectively, with 

\begin{equation}
\label{Lambda}
\Lambda_{s_{1}}=\log\left(\frac{x_{12}x_{13}\sin\left(\phi_{1}\right)}{x_{23}\epsilon\sin\left(\frac{1}{2}\left(\phi_{1}-\phi_{2}+\phi_{3}\right)\right)}\right),\ \ \Lambda_{t_{1}}=\Lambda_{s_{1}}+\delta x_{1},
\end{equation}

and the other cutoffs defined by cyclic permutation of the indices. Normalisation is then obtained by dividing by $\mathcal{N}_{i}=\epsilon^{\Delta^{i}_{0}}\frac{F_{0}^{i}\left(0\right)}{||F_{0}^{i}||}\sqrt{\frac{2}{-\Delta_{0}^{i}}}$, which was obtained in \cite{Cavaglia:2018lxi} so as to ensure canonical normalisation of the two-cusped loop's expectation value to $\frac{1}{|x_{1}-x_{2}|^{2\Delta_{0}}}$.

\section{Scalar insertions at the L cusp}

\begin{figure}
\label{tringle}
\begin{center}
\includegraphics[width=11cm]{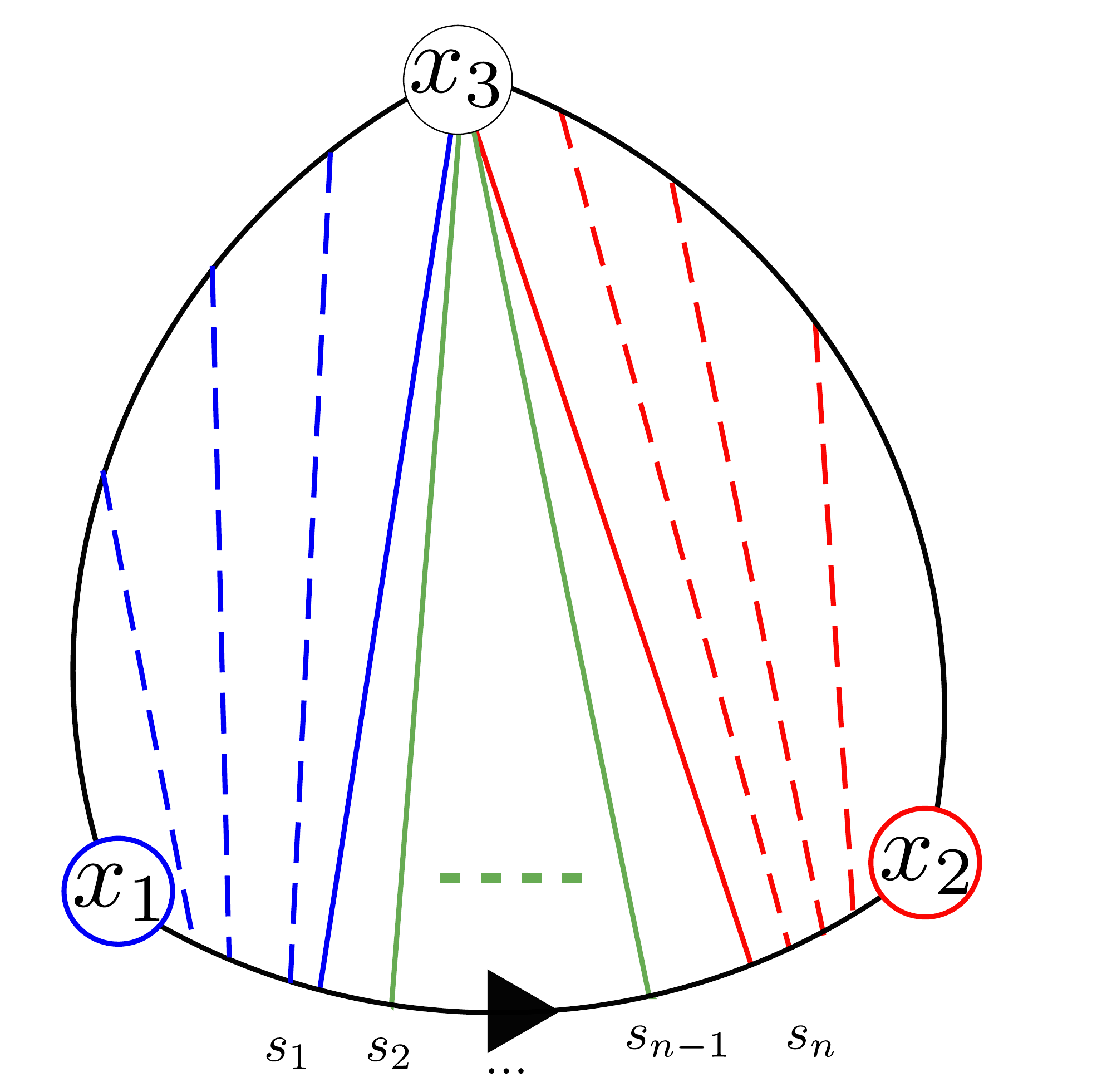}
\end{center}
\caption{Leading order diagram for the cusp with insertions. Blue and red dashed lines are resummed up to the blue and red solid lines respectively, and the solid lines (one red, one blue and $n-2$ green) are the propagators between the light cusp 3 and the arc opposite it.}
\centering
\end{figure}

In this section we carry out the calculation explicitly. The leading order Feynman diagrams relevant to (\ref{insertions}) take the form of $n$ propagators from $x_{3}$ to points along the arc connecting $x_{1}$ and $x_{2}$ and a resummation of ladders around the cusps at $x_{1}$ and $x_{2}$ up to the points where propagators from $x_{3}$ meet the arc (see figure \ref{tringle}).
\\

Apart from the factor $\hat{g}_{1}^{2m_{1}}\cdot\hat{g}_{2}^{2m_{2}}$ which we suppress, resumming all of these diagrams amounts to the following integral :

\begin{equation} \label{integral}
\begin{aligned}[b]
\mathbf{C} = & \frac{1}{|2\frac{d\zeta_{13}\left(0\right)}{ds}|^{n}}\cdot \int_{-\infty}^{0}\frac{ds_{1}}{\cosh\left(s_{1}-\delta x_{1}\right)+\cos\left(\phi_{1}\right)}\cdot G_{1}\left(\Lambda_{t_{1}},\Lambda_{t_{1}},s-\delta x_{1},0\right)\\
& \times \int_{s_{1}}^{0}\frac{ds_{2}}{\cosh\left(s_{2}-\delta x_{1}\right)+\cos\left(\phi_{1}\right)}\ ...\ \int_{s_{n-2}}^{0}\frac{ds_{n-1}}{\cosh\left(s_{n-1}-\delta x_{1}\right)+\cos\left(\phi_{1}\right)}\\ 
& \times \int_{s_{n-1}}^{0}\frac{ds_{n}}{\cosh\left(s_{n}-\delta x_{1}\right)+\cos\left(\phi_{1}\right)}\cdot G_{2}\left(\Lambda_{t_{2}},\Lambda_{t_{2}},-\delta x_{2},T_{12}\left(s_{n}\right)\right). 
\end{aligned}
\end{equation}
\\

Note that although this expression and the informing parametrisation are not symmetric with respect to the points $x_{1}$ and $x_{2}$ our final expression will be. $G_{1}$ and $G_{2}$ are the ladder resummations about cusp $1$ and $2$ as in (\ref{sum}). We have incorporated the necessary shifts $\delta x_{1}$ and $\delta x_{2}$ so as to ensure that our propagators are properly placed, and stress that we have not yet divided by the renormalising factors. Each of the factors under the integration measures come from the propagators, along with the factor sitting outside the first integral (explicitly, these factors are introduced when using (\ref{derivatives}) to replace the primal coordinate-free propagator expressions with those pertaining to the parametrisation). 
\\

In the small $\epsilon$ or large $\Lambda$ limit, the dominant contributions from (\ref{sum}) are the terms containing the ground state function $F_{0}$ multiplied by the leading exponential term in $\sinh$. This can be seen by noting that $G_{1}(\Lambda_{t_{1}},\Lambda_{t_{1}},s-\delta x_{1},0)$ only depends on the cutoff $\Lambda_{t_{1}}$ through the $\sinh$ terms in (\ref{sum}). The fastest growing of these $\sinh$ terms is that in the ground state contribution to the sum. The contributions from bound states of higher energies are slower exponentials in $\Lambda_{t_{1}}$ and the contributions from unbound states are oscillatory. A similar statement holds for $G_{2}\left(\Lambda_{t_{2}},\Lambda_{t_{2}},-\delta x_{2},T_{12}\left(s_{n}\right)\right)$.
\\

Singling these dominant contributions out and collecting prefactors gives

\begin{equation}
\label{moreintegral}
\begin{aligned}[b]
\mathbf{C}=&\frac{4}{|2\frac{d\zeta_{13}\left(0\right)}{ds}|^{n}}\cdot \frac{F_{0}^{1}\left(0\right)F_{0}^{2}\left(0\right)}{||F_{0}^{1}||^{2}||F_{0}^{1}||^{2}\left(-\Delta_{0}^{1}\right)\left(-\Delta_{0}^{2}\right)}\cdot\exp\left(-\Lambda_{t_{1}}\Delta_{0}^{1}-\Lambda_{t_{2}}\Delta_{0}^{2}\right)\\
&\times\int_{-\infty}^{0}\frac{ds_{1}}{\cosh\left(s_{1}-\delta x_{1}\right)+\cos\left(\phi_{1}\right)}\cdot F_{0}^{1}\left(s_{1}-\delta x_{1}\right)\cdot\exp\left(\frac{\Delta_{0}^{1}}{2}\left(\delta x_{1}-s_{1}\right)\right)\\  
&\times \int_{s_{1}}^{0}\frac{ds_{2}}{\cosh\left(s_{2}-\delta x_{1}\right)+\cos\left(\phi_{1}\right)}\ ...\ \int_{s_{n-2}}^{0}\frac{ds_{n-1}}{\cosh\left(s_{n-1}-\delta x_{1}\right)+\cos\left(\phi_{1}\right)}\\
&\times \int_{s_{n-1}}^{0}\frac{ds_{n}}{\cosh\left(s_{n}-\delta x_{1}\right)+\cos\left(\phi_{1}\right)}\cdot F_{0}^{2}\left(T_{12}\left(s_{n}\right)+\delta x_{2}\right)\cdot \exp\left(\frac{\Delta_{0}^{2}}{2}\left(\delta x_{2}-T_{12}\left(s_{n}\right)\right)\right). 
\end{aligned}
\end{equation}
\\

Again, the superscripts on the $F$ pertain to which cusp they are relevant to. The ground state functions $F^{i}_{0}$ are even with $F^{i}_{0}\left(z\right)=F^{i}_{0}\left(-z\right)$. Therefore in addition to (\ref{integraltransform}) there is also another otherwise identical relationship with $z\rightarrow-z$. Using both of these relations to replace the instances of $F$ appearing under the integrals in (\ref{moreintegral}) with Q-functions yields

\begin{equation}
\label{moremoreintegral}
\begin{aligned}[b]
\mathbf{C}=&\frac{-4}{|2\frac{d\zeta_{13}\left(0\right)}{ds}|^{n}}\cdot \frac{F_{0}^{1}\left(0\right)F_{0}^{2}\left(0\right)}{||F_{0}^{1}||^{2}||F_{0}^{1}||^{2}\left(-\Delta_{0}^{1}\right)\left(-\Delta_{0}^{2}\right)}\cdot\exp\left(-\Lambda_{t_{1}}\Delta_{0}^{1}-\Lambda_{t_{2}}\Delta_{0}^{2}\right)\\   
&\times \int_{-\infty}^{0}\frac{ds_{1}}{\cosh\left(s_{1}-\delta x_{1}\right)+\cos\left(\phi_{1}\right)}\cdot\int_{|}q_{0}^{1}\left(u\right)e^{w_{\phi_{1}}\left(\delta x_{1}-s_{1}\right)u}\frac{du}{u}\\   
&\times \int_{s_{1}}^{0}\frac{ds_{2}}{\cosh\left(s_{2}-\delta x_{1}\right)+\cos\left(\phi_{1}\right)}\ ...\ \int_{s_{n-2}}^{0}\frac{ds_{n-1}}{\cosh\left(s_{n-1}-\delta x_{1}\right)+\cos\left(\phi_{1}\right)}\\  
&\times\int_{s_{n-1}}^{0}\frac{ds_{n}}{\cosh\left(s_{n}-\delta x_{1}\right)+\cos\left(\phi_{1}\right)}\cdot\int_{|}q_{0}^{2}\left(v\right)e^{w_{\phi_{2}}\left(-T_{12}\left(s_{n}\right)-\delta x_{2}\right)v}\frac{dv}{v}\cdot e^{\Delta_{0}^{2}\delta x_{2}}.
\end{aligned}
\end{equation}
\\

Things are simplified further by applying the relation between the cutoff parameters $\Lambda_{t_{i}}=\Lambda_{s_{i}}+\delta x_{i}$ in the exponential prefactor. It is at this stage that another one of the strengths of the parametrisation comes into play, in the form of the useful relation

\begin{equation}
\label{surprise}
w_{\phi_{2}}\left(-\delta x_{2}-T_{12}\left(s\right)\right)=w_{\phi_{1}}\left(s-\delta x_{1}\right)-\phi_{3}.
\end{equation}
\\

Using (\ref{surprise}) we can eliminate $T_{12}$ from our expression. This leads to

\begin{equation}
\label{moremoremoreintegral}
\begin{aligned}[b]
\mathbf{C}=&\frac{-4}{|2\frac{d\zeta_{13}\left(0\right)}{ds}|^{n}}\cdot \frac{F_{0}^{1}\left(0\right)F_{0}^{2}\left(0\right)}{||F_{0}^{1}||^{2}||F_{0}^{1}||^{2}\left(-\Delta_{0}^{1}\right)\left(-\Delta_{0}^{2}\right)}\cdot\exp\left(-\Lambda_{t_{1}}\Delta_{0}^{1}-\Lambda_{s_{2}}\Delta_{0}^{2}\right)\\   
&\times \int_{-\infty}^{0}\frac{ds_{1}}{\cosh\left(s_{1}-\delta x_{1}\right)+\cos\left(\phi_{1}\right)}\cdot\int_{|}q_{0}^{1}\left(u\right)e^{w_{\phi_{1}}\left(\delta x_{1}-s_{1}\right)u}\frac{du}{u}\\   
&\times \int_{s_{1}}^{0}\frac{ds_{2}}{\cosh\left(s_{2}-\delta x_{1}\right)+\cos\left(\phi_{1}\right)}\ ...\ \int_{s_{n-2}}^{0}\frac{ds_{n-1}}{\cosh\left(s_{n-1}-\delta x_{1}\right)+\cos\left(\phi_{1}\right)}\\  
&\times\int_{s_{n-1}}^{0}\frac{ds_{n}}{\cosh\left(s_{n}-\delta x_{1}\right)+\cos\left(\phi_{1}\right)}\cdot\int_{|}q_{0}^{2}\left(v\right)e^{w_{\phi_{1}}\left(s_{n}-\delta x_{1}\right)v}e^{-\phi_{3}v}\frac{dv}{v}.
\end{aligned}
\end{equation}
\\

Another grace of the parametrisation is the fact that 

\begin{equation}
\label{wdoesitagain}
w_{\phi_{i}}'\left(z\right)=\frac{-\sin\left(\phi_{i}\right)}{\cosh\left(z\right)+\cos\left(\phi_{i}\right)},  
\end{equation}

which allows us to straightforwardly carry out each integral over $s_{i}$ for $2\leq i\leq n$. Since $w_{\phi_{i}}\left(-\delta x_{1}\right)=-\phi_{2}+\phi_{3}$ the terms that arise from evaluation of the primitives at $s_{i}=0$ are exponentially suppressed in $v$ (bear in mind the triangle inequalities on the $\phi_{i}$) so vanish at $\Re(v)=\infty$. The Q-functions have no poles outside of the imaginary axis so we can move the $v$ contour to $\Re(v)=\infty$ where this vanishing occurs. We therefore need only to retain terms arising from evaluation of the primitives at $s_{i}=-\infty$ with the understanding that we take $\Re(v)\rightarrow\infty$. This gives us

\begin{equation}
\label{moremoremoremoreintegral}
\begin{aligned}[b]
\mathbf{C}=&\frac{-4}{|2\frac{d\zeta_{13}\left(0\right)}{ds}|^{n}}\cdot \frac{F_{0}^{1}\left(0\right)F_{0}^{2}\left(0\right)}{||F_{0}^{1}||^{2}||F_{0}^{1}||^{2}\left(-\Delta_{0}^{1}\right)\left(-\Delta_{0}^{2}\right)}\cdot\exp\left(-\Lambda_{t_{1}}\Delta_{0}^{1}-\Lambda_{s_{2}}\Delta_{0}^{2}\right)\\   
&\times \int_{-\infty}^{0}\frac{ds_{1}}{\cosh\left(s_{1}-\delta x_{1}\right)+\cos\left(\phi_{1}\right)}\cdot\int_{|}q_{0}^{1}\left(u\right)e^{w_{\phi_{1}}\left(\delta x_{1}-s_{1}\right)u}\frac{du}{u}\\   
&\times \left(\frac{1}{\sin\left(\phi_{1}\right)}\right)^{n-1}\int_{|}q_{0}^{2}\left(v\right)e^{w_{\phi_{1}}\left(s_{1}-\delta x_{1}\right)v}e^{-\phi_{3}v}\frac{dv}{v^n}.
\end{aligned}
\end{equation}
\\

Note that $w_{\phi}(z)$ is odd and so the integral over $s_{1}$ is carried out in much the same way except that a factor of $\frac{1}{u-v}$ arises rather than simply $\frac{1}{v}$. As $z\rightarrow\infty$, $w_{\phi_{i}}\left(z\right)\rightarrow\phi_{i}$ and so

\begin{equation}
\label{moremoremoremoremoreintegral}
\begin{aligned}[b]
\mathbf{C}=&\frac{-4}{|2\frac{d\zeta_{13}\left(0\right)}{ds}|^{n}}\cdot \frac{F_{0}^{1}\left(0\right)F_{0}^{2}\left(0\right)}{||F_{0}^{1}||^{2}||F_{0}^{1}||^{2}\left(-\Delta_{0}^{1}\right)\left(-\Delta_{0}^{2}\right)}\cdot\exp\left(-\Lambda_{t_{1}}\Delta_{0}^{1}-\Lambda_{s_{2}}\Delta_{0}^{2}\right)\\
&\times\left(\frac{1}{\sin\left(\phi_{1}\right)}\right)^{n}\cdot\int_{|}\frac{du}{u}\int_{|}\frac{dv}{v^n}\cdot q_{0}^{1}\left(u\right)\cdot q_{0}^{2}\left(v\right)\frac{e^{-\phi_{3}v}}{u-v}\left(e^{\left(-\phi_{2}+\phi_{3}\right)\left(v-u\right)}-e^{\phi_{1}\left(v-u\right)}\right). 
\end{aligned}
\end{equation}
\\

The integrand actually has no pole at $u=v$, and so the $u$ and $v$ contours can be moved over each other without concern. We therefore fix the $v$ contour to be to the right of the $u$ contour. \\

Additionally, the integral with integrand proportional to $e^{-\phi_{2}v}$ can then be discarded, as we are taking the $v$ contour to $\Re\left(v\right)=\infty$ where this integrand vanishes. The remaining integral over u is then straightforwardly carried out by closing the u contour with a semicircle in the right half of the complex plane and picking up the residue at $u$ = $v$:

\begin{equation}
\label{homestretch}
\begin{aligned}[b]
\mathbf{C}=&\frac{-4}{|2\frac{d\zeta_{13}\left(0\right)}{ds}|^{n}}\cdot \frac{F_{0}^{1}\left(0\right)F_{0}^{2}\left(0\right)}{||F_{0}^{1}||^{2}||F_{0}^{1}||^{2}\left(-\Delta_{0}^{1}\right)\left(-\Delta_{0}^{2}\right)}\cdot\exp\left(-\Lambda_{t_{1}}\Delta_{0}^{1}-\Lambda_{s_{2}}\Delta_{0}^{2}\right)\\
&\times-2\pi i\cdot\left(\frac{1}{\sin\left(\phi_{1}\right)}\right)^{n}\cdot\int_{|}\frac{dv}{v^{n+1}}\cdot q_{0}^{1}\left(v\right)\cdot q_{0}^{2}\left(v\right)e^{-\phi_{3}v}.  
\end{aligned}
\end{equation}
\\

We have here (from differentiation of the coordinate functions as in (\ref{arcs})) the first sighting of our correlator's conformal dependence on $x_{3}$ in this analysis:

\begin{equation}
\label{boring}
|2\frac{d\zeta_{13}\left(0\right)}{ds}|=\frac{2|z_{1}-z_{3}||z_{2}-z_{3}|\sin\left(\frac{\phi_{1}+\phi_{2}-\phi_{3}}{2}\right)}{|z_{1}-z_{2}|\sin\left(\phi_{1}\right)}.
\end{equation}
\\

Also, we should note this relation for the norm of the wavefunctions from \cite{Cavaglia:2018lxi}:

\begin{equation}
\label{norm}
||F_{0}^{k}||=2\pi i\cdot\sqrt{\frac{2}{\Delta_{0}^{k}}}\sqrt{\int_{|}\frac{\left(q_{0}^{k}\left(u\right)\right)^{2}}{u}\frac{du}{2\pi i}}. 
\end{equation}
\\

Tedious but straightforward algebra (including applying (\ref{Lambda}) and dividing by the two normalisation constants $\mathcal{N}_{1}$ and $\mathcal{N}_{2}$) then leads to 

\begin{equation}
\label{thing}
\begin{aligned}[b]
\mathbf{C}_{Renormalised}=& \left(\frac{|z_{1}-z_{2}|}{|z_{1}-z_{3}||z_{2}-z_{3}|}\right)^{n}\cdot\left(\frac{|z_{1}-z_{3}|}{|z_{2}-z_{3}||z_{1}-z_{2}|}\right)^{\Delta_{0}^{2}}\cdot\left(\frac{|z_{2}-z_{3}|}{|z_{1}-z_{2}||z_{1}-z_{3}|}\right)^{\Delta_{0}^{1}}\\
&\times\left(2\sin\left(\frac{\phi_{1}+\phi_{2}-\phi_{3}}{2}\right)\right)^{-n+\Delta_{0}^{1}+\Delta_{0}^{2}}\cdot\left(2\sin\left(\phi_{1}\right)\right)^{-\Delta_{0}^{1}}\cdot\left(2\sin\left(\phi_{2}\right)\right)^{-\Delta_{0}^{2}}\\   &\times\frac{2\cdot2\pi i}{||F_{0}^{1}||||F_{0}^{2}||\sqrt{\left(-\Delta_{0}^{1}\right)\left(-\Delta_{0}^{2}\right)}}\cdot\int_{|}\frac{dv}{v^{n+1}}\cdot q_{0}^{1}\left(v\right)\cdot q_{0}^{2}\left(v\right)e^{-\phi_{3}v}\\
\; &\; \\
=&\; \hat{\mathbf{C}}\cdot\left(\frac{|z_{1}-z_{2}|}{|z_{1}-z_{3}||z_{2}-z_{3}|}\right)^{n}\cdot\left(\frac{|z_{1}-z_{3}|}{|z_{2}-z_{3}||z_{1}-z_{3}|}\right)^{\Delta_{0}^{2}}\cdot\left(\frac{|z_{2}-z_{3}|}{|z_{2}-z_{3}||z_{1}-z_{3}|}\right)^{\Delta_{0}^{1}}.
\end{aligned}
\end{equation}
\\

As it should, this has the form of a 3-point correlator in a conformal field theory. It should again be stated that we have suppressed the coupling factors $\hat{g}_{1}^{2m_{1}}\cdot \hat{g}_{2}^{2m_{2}}$. One can replace the $||F_{0}^{i}||$ in the above using (\ref{norm}). The Q-functions behave as $u^{\Delta}e^{\phi u}$ for large u, and so the structure constant $\hat{\mathbf{C}}$ is further simplified through use of the bracket (\ref{bracket}) which encapsulates the extraneous factors:

\begin{equation}
\label{yay}
\boxed{\hat{\mathbf{C}}=\frac{\langle q_{0}^{1}\left(u\right)q_{0}^{2}\left(u\right)\frac{e^{-\phi_{3}u}}{u^{n}}\rangle}{\sqrt{\langle\left(q_{0}^{1}\right)^{2} \rangle}\sqrt{\langle\left(q_{0}^{2}\right)^{2} \rangle}} .}
\end{equation}
\\

Note that this entire analysis can be repeated with each of the H cusps excited to level $k$ and $l$, and the result therein amounts to replacing each Q-function $q^{i}_{0}$ in (\ref{yay}) with the corresponding excited Q-functions $q^{1}_{k}$ and $q^{2}_{l}$ as in (\ref{integraltransform}). It should be born in mind that one also needs to use a different normalisation for these excited cusps as laid out in \cite{Cavaglia:2018lxi}, but the form of the result is the same as (\ref{yay}) when using the bracket notation.
\\

\section{Variation of \texorpdfstring{$\Delta$}{Lg} and Lagrangian insertions}\label{derivativesection}

This section considers the vacuum expectation value of the two-cusped Wilson loop (in the fundamental representation) which has scaling dimension $\Delta$:

\begin{equation}
\label{2cuspvev}
\langle \frac{1}{N}\Tr W_{x}^{y}(\vec{n}_{1}) W_{y}^{x}(\vec{n}_{2}) \rangle =\frac{1}{|x-y|^{2\Delta}}.
\end{equation}

Note that only a single Q-function is relevant to this object, which we denote $q$. There is similarly only a single coupling to consider, $\hat{g}^2=g^{2}\cdot\frac{\cos(\theta)}{2}$ with $\vec{n}_{1}\cdot\vec{n}_{2}=\cos(\theta)$. In \cite{Cavaglia:2018lxi} perturbation theory was applied to find the variation of $\Delta$ with $\hat{g}^2$:

\begin{equation}
\label{derivativedelta}
-\frac{1}{4}\frac{\partial\Delta}{\partial\hat{g}^{2}}=\frac{\langle{q^2}\frac{1}{u}\rangle}{\langle q^2\rangle}.
\end{equation}

As was stated there, this derivative can be interpreted as the structure constant for the correlator between the two cusps and the Lagrangian \cite{Costa:2010rz}. As we shall show, the diagrams relevant to the computation of this correlator in fact bear strong resemblances to those considered earlier in this paper. 
\\

\begin{figure}
    \centering
    \includegraphics[width=14cm]{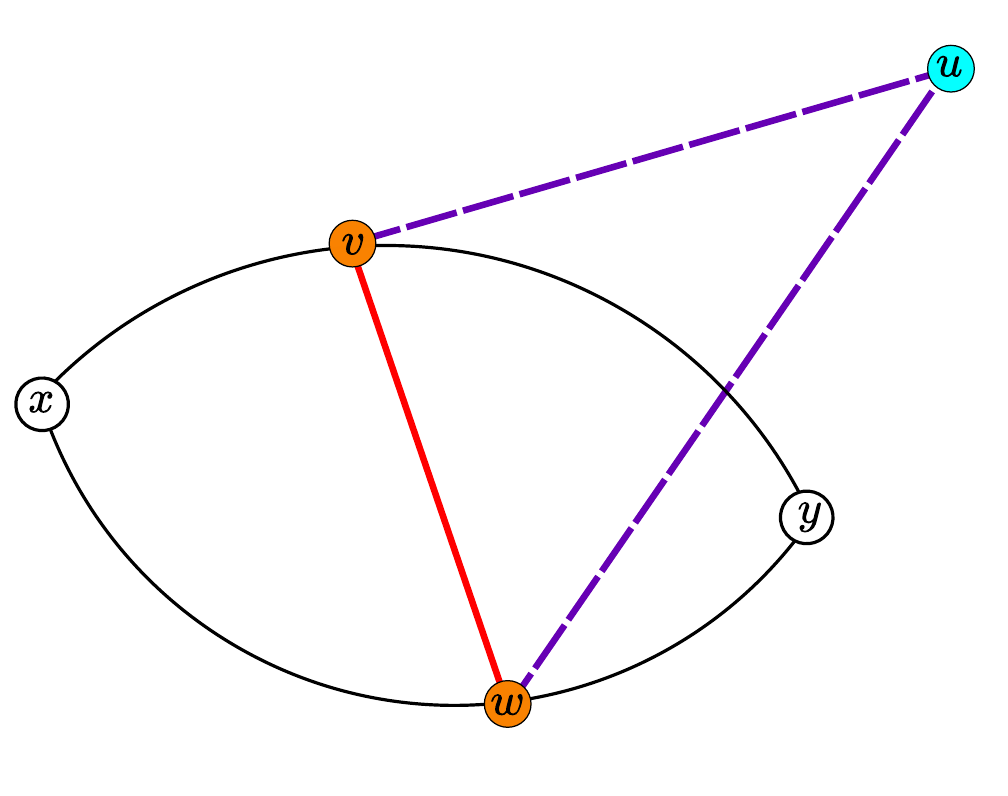}
    \caption{Propagators between the Lagrangian at $u$ and scalar fields at points $v$, $w$ on the loop are drawn in purple. Integrating over $u$ gives rise to an effective propagator between the arcs drawn in red.}
    \label{lagrangianpicture}
\end{figure}

First, consider how the vacuum expectation value (\ref{2cuspvev}) varies with the coupling $\hat{g}^2$:

\begin{equation}
\label{cderivatives}
\frac{\partial}{\partial\hat{g}^{2}} \frac{1}{|x-y|^{2\Delta}}= \frac{-2\log(\Upsilon|x-y|)}{|x-y|^{2\Delta}}\cdot \frac{\partial\Delta}{\partial\hat{g}^{2}}
\end{equation}

where $\Upsilon$ is a cutoff. The relevance to a correlator with a Lagrangian is manifested when we differentiate the path integral as below. Recall here that $\hat{g}^2=g^2\cdot\frac{\cos(\theta)}{2}=g_{SYM}^{2}\cdot N\cdot\frac{\cos(\theta)}{32\pi^2}$.

\begin{equation}
\label{pathint}
\begin{aligned}[b]
\frac{\partial}{\partial\hat{g}^{2}} \int\mathcal{D}[\Phi]&\exp\left[\frac{-1}{g_{SYM}^2}\int d^{4}q\mathcal{L}(q)\right]\frac{1}{N}\Tr W_{x}^{y}(\vec{n}_{1}) W_{y}^{x}(\vec{n}_{2})\\
&\\
&=\frac{32\pi^2}{N^2\cdot g_{SYM}^4\cdot\cos(\theta)}\int\mathcal{D}[\Phi]\exp\left[\frac{-1}{g_{SYM}^2}\int d^{q}v\mathcal{L}(q)\right]\left(\int d^{4}u\mathcal{L}(u)\right)\Tr W_{x}^{y}(\vec{n}_{1}) W_{y}^{x}(\vec{n}_{2})\\
&=\frac{32\pi^2}{N^2\cdot g_{SYM}^4\cdot\cos(\theta)}\langle\left(\int d^{4}u\mathcal{L}(u)\right)\Tr W_{x}^{y}(\vec{n}_{1}) W_{y}^{x}(\vec{n}_{2})\rangle\\
&=\frac{32\pi^2}{N^2\cdot g_{SYM}^4\cdot\cos(\theta)}\int d^{4}u\ \langle\mathcal{L}(u)\Tr W_{x}^{y}(\vec{n}_{1}) W_{y}^{x}(\vec{n}_{2})\rangle
\end{aligned}
\end{equation}
\\

where we work in Euclidean signature. The full four-dimensional $\mathcal{N}=4$ Lagrangian notoriously contains a number of fields and interactions but in the ladders limit the prefactor of $(\cos(\theta))^{-1}$ will suppress almost all of these. The only surviving terms will be a subset of those due to the $\Phi$ kinetic term, $\Tr\left(\partial_{\mu}\Phi^{c}\partial^{\mu}\Phi^{c}\right)$. The relevant diagrams are those where a $\Phi$ propagator joins the inserted Lagrangian to each of the Wilson loop's arcs, as in figure \ref{lagrangianpicture}.
\\

We are led to analyse the subcorrelator between the $\Tr\left(\partial_{\mu}\Phi^{c}\partial^{\mu}\Phi^{c}\right)$ inserted at $u$ and the scalar combinations $n_{1}^{a}\Phi^{a}$ and $n_{2}^{b}\Phi^{b}$ inserted at $v$ and $w$ respectively. Integrating this quantity over $u$ will define an effective propagator between the scalars at $v$ and $w$.
\\

This subcorrelator is $\langle\Tr\left(\partial_{\mu}\Phi^{i}(u)\partial^{\mu}\Phi^{i}(u)\right)\Tr\left(n_{1}^{a}\Phi^{a}(v)n_{2}^{b}\Phi^{b}(w)\right)\rangle$. The two traces supply a factor of $\frac{1}{4}$ and there are two ways of contracting the fields. With the derivatives applied to the propagators we find that this equals 

\begin{equation}
\label{subc}
\frac{2N^2\cdot g_{SYM}^4\cos(\theta)}{(4\pi^2)^2}\cdot\frac{(u^{\mu}-v^{\mu})(u_{\mu}-w_{\mu})}{|u-v|^4|u-w|^4}.
\end{equation}

We only retain the leading $N^2$ term. The integral of this function over $u\in\mathbb{R}^4$ is found to be $\frac{N^2\cdot g_{SYM}^4\cos(\theta)}{8\pi^2}\cdot\frac{1}{|v-w|^2}$. There is one such contribution for each choice of $v$ and $w$ on the arcs together with the ladder resummations. Denote by $L_{x}(v,w)$ and $L_{y}(v,w)$ the sum of ladder diagrams centred on the cusps at $x$ and $y$ respectively) up to the points $v$ and $w$ on the arcs.   (\ref{pathint}) then equals

\begin{equation}
\label{familiar}
4\int dv \int dw\  \frac{1}{|v-w|^2} L_{x}(v,w) L_{y}(v,w)
\end{equation}

with the $v$ and $w$ integral along the upper and lower arc respectively. The point of this section is that for a fixed $v$ the $w$ integral is the same as for a single scalar insertion in the HHL loop with $\phi_{3}=0$ and no change in $\vec{n}$ across the L cusp. We transplant that result here. There then remains a single integral along the upper arc to perform:
\\

\begin{equation}
\label{lastintegral}
\frac{\langle{q^2}\frac{1}{u}\rangle}{\langle q^2\rangle}\cdot \frac{4}{|x-y|^{2\Delta}} \cdot\int dv\ \frac{|x-y|}{|v-x||v-y|}.
\end{equation}
\\

This is the source of the logarithmic divergence which we deal with via the cutoff $\Upsilon$. Specifically, we integrate over the arc to a distance $\Upsilon^{-1}$ from $x$ and $y$. This is straightforward and we arrive at

\begin{equation}
\label{lastthing}
\frac{\partial}{\partial\hat{g}^{2}} \frac{1}{|x-y|^{2\Delta}}=\frac{8\log(\Upsilon|x-y|)}{|x-y|^{2\Delta}}\cdot\frac{\langle{q^2}\frac{1}{u}\rangle}{\langle q^2\rangle}.
\end{equation}

Comparison of this result with (\ref{cderivatives}) gives another proof of (\ref{derivativedelta}).

\section{Perspectives on projectors}\label{perspectives}

As has been stated, at large $\Lambda$ the sum in (\ref{sum}) is dominated by the lower energy terms. We write here the asymptotic expression for the all loop ladder resummation between the four marked points on the 2-cusped loop in figure \ref{2cusp}, written using $\Delta_{k}=-\sqrt{-E_{k}}$ and retaining only the bound states (of which any number exist at sufficiently large coupling):

\begin{equation}
\label{sum2}
G\left(\Lambda_{1},\Lambda_{2},\Lambda_{3},\Lambda_{4}\right)\simeq\sum_{k}\frac{2F_{k}\left(\Lambda_{1}-\Lambda_{2}\right)F_{k}\left(\Lambda_{4}-\Lambda_{3}\right)}{||F_{k}||^{2}(-\Delta_{k})}\exp\left(\frac{-\Delta_{k}}{2}\left(\Lambda_{1}+\Lambda_{2}+\Lambda_{3}+\Lambda_{4}\right)\right).
\end{equation}

This is saturated by the ground state contribution. One could act on $G$ by the following operators $\mathcal{O}_{n}$ which eliminate the states of energy up to $E_{n}$:

\begin{equation}
\label{proj}
\mathcal{O}_{2m}=\prod_{i=0}^{m-1}\frac{\partial_{+}+\Delta_{2i}}{-\Delta_{2m}+\Delta_{2i}}, \ \ \ 
\mathcal{O}_{2m+1}=\prod_{i-0}^{m-1}\frac{\partial_{+}+\Delta_{2i+1}}{-\Delta_{2m+1}+\Delta_{2i+1}}\times\partial_{-}
\end{equation}

in which $\partial_{\pm}\equiv\partial_{\Lambda_{1}}\pm\partial_{\Lambda_{2}}$. One also has $\bar{\partial}_{\pm}\equiv\partial_{\Lambda_{4}}\pm\partial_{\Lambda_{3}}$, with operators $\Bar{\mathcal{O}}_{n}$ defined by (\ref{proj}) but with $\partial\rightarrow\Bar{\partial}$.
\\

These projection operators were first given in section 5 of \cite{Cavaglia:2018lxi}. There they were used to define $W_{n}\equiv\mathcal{O}_{n}\bar{\mathcal{O}}_{n}G(\Lambda_{1},\Lambda_{2},\Lambda_{3},\Lambda_{4})\rvert_{\Lambda_{1}=\Lambda_{2}=\Lambda_{3}=\Lambda_{4}=\Lambda}$. In the limit of large $\Lambda$ this quantity was found to have the structure of a two-point correlator of operators with dimension $\Delta_{n}$.
\\

These two-point correlators can be seen to be the two-point functions of the cusps with insertions produced by the $\mathcal{O}_{n}$. The exact form of the insertions is sensitive to the regularisation scheme imposed but in the point-splitting regularisation the operator $\mathcal{O}_{1}=\frac{1}{2}\partial_{-}$ acts on (\ref{loop}) to produce an insertion at cusp 3 proportional to $(\Phi^{a}n_{3,1}^{a}-\Phi^{a}n_{3,2}^{a})$, which is exactly the kind of insertion we study here. Taking further derivatives will either produce additional such insertions by action on the exponential in the Wilson loop's definition or will produce insertions containing derivatives by action on previous insertions.
\\

This procedure allows for two pictures concerning the menagerie of insertions produced by the projections. In one, the methods of \cite{Cavaglia:2018lxi} allow for an expression in terms of excited Q functions. In the other, evaluating integrals as done in this paper gives a different expression containing functions of $u$ that are analogous to the $u^{-n}$ that appears with $n$ scalar insertions.
\\

The $n^{th}$ order projection operator produces insertions that are finite sums of various products of $\Phi$ and its derivatives. In the couplings $\hat{g_{i}}$ the highest order of these are of the type studied in this paper, a product of $n$ scalars. The lowest order terms (after the constants produced at odd orders) are $n^{th}$ derivatives of $\Phi$. 
\\

\section{Conclusions and discussion}


We supply a novel set of vacuum expectation values computed to all orders in perturbation theory in the ladders limit. This supports the use of the Quantum Spectral Curve in approaching cusped Wilson loops. 
\\

Equation (1.5) in \cite{Cavaglia:2018lxi} gives a simple expression for the derivative of the cusp anomalous dimension $\Delta$ with respect to the square of the coupling $\hat{g}$, which was interpreted as the structure constant of two cusps with a single BPS operator. Our result gives agreement with this by enabling a parallel derivation.
\\

There is a more general class of insertions to consider. These consist of derivatives of the scalar fields of various orders. Schematically these are $\phi '$, $\phi ''$, et cetera and various products of these. Combinations may include such things as $(\phi ')^2 \phi '''$, $(\phi)^3 (\phi ')^2$, $\phi''\phi^{721077}$, $\phi (\phi ')^3 (\phi '')^2$, and so on. The projection operators defined in \cite{Cavaglia:2018lxi} that excite a cusp act so as to produce specific sums of such insertions which justifies some restriction of attention.
\\

What could be useful is a general expression for the functions $P(u)$ that appear inside the bracket alongside $q_{1}$ and $q_{2}$ when one computes the vacuum expectation values of correlators with the insertions produced by the $k^{th}$ projection operators. One obstacle to this is that some modification of the parametrisation of the arcs is required to deal with the derivatives of the propagators.
\\

With or without such a formula, one could still hope to make comparisons between expressions found in this way and those given in \cite{Cavaglia:2018lxi} for the HLL cusp with a single L cusp at $k^{th}$ excitation. There is a belief that the Q-functions at zero coupling and $k^{th}$ excitation should be related to the $k^{th}$ $P(u)$ by some involutory transformation. Progress towards an expression for the HHH structure constant, which remains an interesting open problem, may be achieved in this way.
\\

This could progress through study of the HLL loop with insertions. There would be one expression for this object's VEV in terms of $k$-excited Q-functions at zero coupling as given in \cite{Cavaglia:2018lxi} and a complementary one involving the as yet undetermined $k^{th}$ $P(u)$.  The relation between these two expressions should give a transformation taking these zero-coupling Q-functions to the $P(u)$. Applying these transformations to the Q-functions at nonzero coupling should give expressions for the HHL loop with insertions at an H cusp. These results would provide several examples of loops with matching insertions at a cusp --- in half of these the cusp is light and in the other half heavy. This may shed light on how to approach the HHH case by offering some insight into the procedure of replacing an L cusp by an H cusp.

\acknowledgments

We are very grateful to A. Cavaglià for helpful discussion and useful advice on the manuscript. We also owe much thanks to F. Levkovich-Maslyuk for technical comments on the manuscript. Significant gratitude is due to A. Sever and T. Yahav for insightful comments and discussion on related work. At every stage of this work ideas and guidance from N. Gromov proved invaluable.

\providecommand{\href}[2]{#2}\begingroup\raggedright\endgroup

\end{document}